\documentclass[a4paper]{jpconf}
\usepackage{graphicx}
\usepackage{wrapfig}

\begin{document}
\title{The CALorimetric Electron Telescope (CALET)
on the International Space Station:
Results from the First Two Years 
On Orbit 
}

\author{
Y~Asaoka$^{1,2}$, O~Adriani$^{3,4}$, Y~Akaike$^{5,6}$, K~Asano$^{7}$, 
M~G~Bagliesi$^{8,9}$, E~Berti$^{3,4}$, G~Bigongiari$^{8,9}$, 
W~R~Binns$^{10}$, S~Bonechi$^{8,9}$, M~Bongi$^{3,4}$, A~Bruno$^{11}$, 
P~Brogi$^{8,9}$, J~H~Buckley$^{10}$, N~Cannady$^{12}$, 
G~Castellini$^{13}$, C~Checchia$^{14,15}$, M~L~Cherry$^{12}$, 
G~Collazuol$^{14,15}$, V~Di~Felice$^{16,17}$, K~Ebisawa$^{18}$, 
H~Fuke$^{18}$, T~G~Guzik$^{12}$, T~Hams$^{5,19}$, N~Hasebe$^{1}$, 
K~Hibino$^{20}$, M~Ichimura$^{21}$, K~Ioka$^{22}$, W~Ishizaki$^{7}$, 
M~H~Israel$^{10}$, K~Kasahara$^{1}$, J~Kataoka$^{1}$, 
R~Kataoka$^{23}$, Y~Katayose$^{24}$, C~Kato$^{25}$, N~Kawanaka$^{26,27}$, 
Y~Kawakubo$^{28}$, K~Kohri$^{29}$, H~S~Krawczynski$^{10}$, 
J~F~Krizmanic$^{19,5}$, T~Lomtadze$^{9}$, P~Maestro$^{8,9}$, 
P~S~Marrocchesi$^{8,9}$, A~M~Messineo$^{30,9}$, J~W~Mitchell$^{6}$, 
S~Miyake$^{31}$, A~A~Moiseev$^{32,19}$, K~Mori$^{1,18}$, M~Mori$^{33}$, 
N~Mori$^{4}$, H~M~Motz$^{34}$, K~Munakata$^{25}$, H~Murakami$^{1}$, 
S~Nakahira$^{35}$, J~Nishimura$^{18}$, G~A~De~Nolfo$^{11}$, 
S~Okuno$^{20}$, J~F~Ormes$^{36}$, S~Ozawa$^{1}$, L~Pacini$^{3,13,4}$, 
F~Palma$^{16,17}$, V~Pal'shin$^{28}$, P~Papini$^{4}$, 
A~V~Penacchioni$^{8,37}$, B~F~Rauch$^{10}$, S~B~Ricciarini$^{13,4}$, 
K~Sakai$^{19,5}$, T~Sakamoto$^{28}$, M~Sasaki$^{19,32}$, Y~Shimizu$^{20}$, 
A~Shiomi$^{38}$, R~Sparvoli$^{16,17}$, P~Spillantini$^{3}$, 
F~Stolzi$^{8,9}$, S~Sugita$^{28}$, J~E~Suh$^{8,9}$, A~Sulaj$^{8,9}$, 
I~Takahashi$^{39}$, M~Takayanagi$^{18}$, M~Takita$^{7}$, T~Tamura$^{20}$, 
N~Tateyama$^{20}$, T~Terasawa$^{35}$, H~Tomida$^{18}$, S~Torii$^{1,40}$, 
Y~Tsunesada$^{41}$, Y~Uchihori$^{42}$, S~Ueno$^{18}$, E~Vannuccini$^{4}$, 
J~P~Wefel$^{12}$, K~Yamaoka$^{43}$, S~Yanagita$^{44}$, A~Yoshida$^{28}$, 
and K~Yoshida$^{45}$
}

\address{
$^{1}$ Waseda Research Institute for Science and Engineering, Waseda University, Japan
}
\address{
$^{2}$ JEM Utilization Center, Human Spaceflight Technology Directorate, \\
\hspace{0.14cm} Japan Aerospace Exploration Agency, Japan
}
\address{
$^{3}$ Department of Physics, University of Florence, Italy
}
\address{
$^{4}$ INFN Sezione di Florence, Italy
}
\address{
$^{5}$ Department of Physics, University of Maryland, USA
}
\address{
$^{6}$ Astroparticle Physics Laboratory, NASA/GSFC, USA
}
\address{
$^{7}$ Institute for Cosmic Ray Research, The University of Tokyo, Japan
}
\address{
$^{8}$ Department of Physical Sciences, Earth and Environment, University of Siena, Italy
}
\address{
$^{9}$ INFN Sezione di Pisa, Italy
}
\address{
$^{10}$ Department of Physics, Washington University, USA
}
\address{
$^{11}$ Heliospheric Physics Laboratory, NASA/GSFC, USA
}
\address{
$^{12}$ Department of Physics and Astronomy, Louisiana State University, USA
}
\address{
$^{13}$ Institute of Applied Physics (IFAC),  National Research Council (CNR), Italy
}
\address{
$^{14}$ Department of Physics and Astronomy, University of Padova, Italy
}
\address{
$^{15}$ INFN Sezione di Padova, Italy
}
\address{
$^{16}$ University of Rome ``Tor Vergata'', Italy
}
\address{
$^{17}$ INFN Sezione di Rome ``Tor Vergata'', Italy
}
\address{
$^{18}$ Institute of Space and Astronautical Science, Japan Aerospace Exploration Agency, Japan
}
\address{
$^{19}$ CRESST and Astroparticle Physics Laboratory NASA/GSFC, USA
}
\address{
$^{20}$ Kanagawa University, Japan
}
\address{
$^{21}$ Faculty of Science and Technology, Graduate School of Science and Technology, \\
\hspace{0.25cm} Hirosaki University, Japan
}
\address{
$^{22}$ Yukawa Institute for Theoretical Physics, Kyoto University, Japan
}
\address{
$^{23}$ National Institute of Polar Research, Japan
}
\address{
$^{24}$ Faculty of Engineering, Division of Intelligent Systems Engineering, \\
\hspace{0.25cm} Yokohama National University, Japan
}
\address{
$^{25}$ Faculty of Science, Shinshu University, Japan
}
\address{
$^{26}$ Hakubi Center, Kyoto University, Japan
}
\address{
$^{27}$ Department of Astronomy, Graduate School of Science, Kyoto University, Japan
}
\address{
$^{28}$ College of Science and Engineering, Department of Physics and Mathematics, \\
\hspace{0.25cm} Aoyama Gakuin University, Japan
}
\address{
$^{29}$ High Energy Accelerator Research Organization, Japan
}
\address{
$^{30}$ University of Pisa, Italy
}
\address{
$^{31}$ Department of Electrical and Electronic Systems Engineering, \\
\hspace{0.25cm} National Institute of Technology, Ibaraki College, Japan
}
\address{
$^{32}$ Department of Astronomy, University of Maryland, College Park, Maryland 20742, USA 
}
\address{
$^{33}$ Department of Physical Sciences, College of Science and Engineering, \\
\hspace{0.25cm} Ritsumeikan University, Japan
}
\address{
$^{34}$ Global Center for Science and Engineering, Waseda University, Japan
}
\address{
$^{35}$ RIKEN, Japan
}
\address{
$^{36}$ Department of Physics and Astronomy, University of Denver, USA
}
\address{
$^{37}$ ASI Science Data Center (ASDC), Italy
}
\address{
$^{38}$ College of Industrial Technology, Nihon University, Japan
}
\address{
$^{39}$ Kavli Institute for the Physics and Mathematics of the Universe,\\
\hspace{0.25cm} The University of Tokyo, Japan
}
\address{
$^{40}$ School of Advanced Science and Engineering, Waseda University, Japan
}
\address{
$^{41}$ Division of Mathematics and Physics, Graduate School of Science, \\
\hspace{0.25cm} Osaka City University, Japan
}
\address{
$^{42}$ National Institutes for Quantum and Radiation Science and Technology, JAPAN
}
\address{
$^{43}$ Nagoya University, Japan
}
\address{
$^{44}$ College of Science, Ibaraki University, Japan
}
\address{
$^{45}$ Department of Electronic Information Systems, Shibaura Institute of Technology, Japan
}

\ead{yoichi.asaoka@aoni.waseda.jp}

\begin{abstract} 
The CALorimetric Electron Telescope (CALET) is a high-energy astroparticle physics space experiment installed on the International Space Station (ISS),  developed and operated by Japan in collaboration with Italy and the United States. 
The CALET mission goals include the investigation of possible nearby sources of high-energy electrons, of the details of galactic particle acceleration and propagation, and of potential signatures of dark matter. CALET measures the cosmic-ray electron$+$positron flux up to 20~TeV, gamma-rays up to 10~TeV, and nuclei with Z=1 to 40 up to 1,000~TeV for the more abundant elements 
during a long-term observation aboard the ISS.
Starting science operation in mid-October 2015, CALET performed continuous observation without major interruption with close to 20 million triggered events over 10 GeV per month. Based on the data taken during the first two-years, we present an overview of CALET observations: 
1) Electron+positron energy spectrum, 2) Nuclei analysis, 3) Gamma-ray observation including a 
characterization of on-orbit performance. Results of the electromagnetic counterpart search
for LIGO/Virgo gravitational wave events are discussed as well.
\end{abstract}
\vspace*{-0.3cm}

\section{Introduction}
The CALorimetric Electron Telescope (CALET)~\cite{torii2015, torii2017} is a high-energy astroparticle physics mission on the ISS, with development and operation conducted by Japan in collaboration with Italy 
and the United States.
The detector was launched to orbit aboard the unmanned H2 
\begin{wrapfigure}{r}{7cm}
\vspace*{-0.2cm}
\begin{center}
\includegraphics[bb=0 0 350 275, width=\hsize]{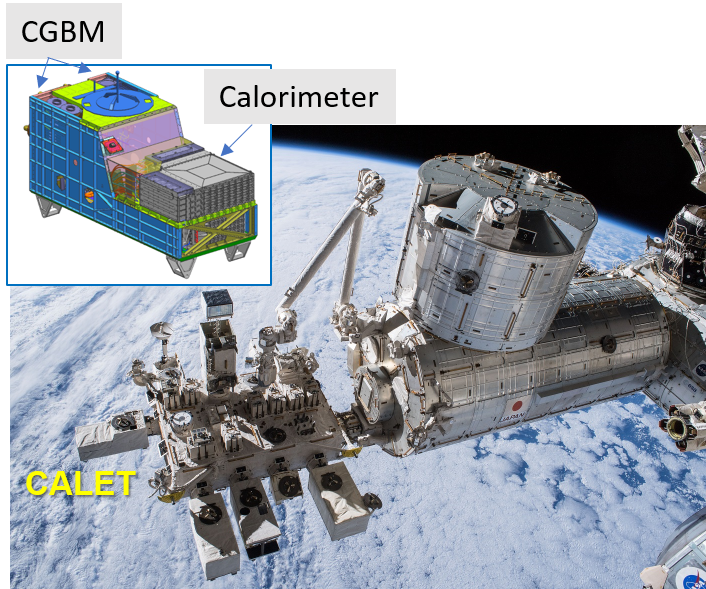}
\end{center}
\vspace*{-0.5cm}
\caption{
JEM-EF and 
the CALET payload attached at the port \#9.
The inset is the CALET instrument package showing the
main calorimeter 
and CALET Gamma-ray Burst Monitor (CGBM) subsystems~\cite{CGBM2013}.
}
\label{fig:CALET}
\vspace*{-0.5cm}
\end{wrapfigure}
Transfer Vehicle (HTV) atop the Japanese H2-B carrier rocket on August 19, 2015.  After arrival of the HTV at the ISS, it was installed on the Japanese Experiment Module--Exposed Facility (JEM-EF). The initial mission duration was two years, extendable to five years or more. The image in Fig.~\ref{fig:CALET} shows the JEM with CALET attached at port \#9 of the JEM-EF, which features a mostly unobstructed field-of-view of 45$^o$ from zenith. The schematic drawing in 
Fig.~\ref{fig:CALET} gives an overview 
of the CALET payload. 

CALET's main instrument is a very thick calorimeter incorporating imaging and total absorption calorimeters. The overall thickness of CALET for normal incidence angle is 30~radiation length, corresponding to $\sim$1.3 proton interaction length. 
Installed on the ISS, CALET carries out long term observation with 
a large area detector, 
thus providing a high-statistics measurement. 
CALET is designed to discover signatures of nearby cosmic-ray accelerators and potentially dark matter in the all-electron (electron$+$positron) spectrum, which is measured with high precision over a wide energy range from 1~GeV to 20~TeV, as well as in the gamma-ray spectrum measured up to 10~TeV. 
Protons, helium, and heavier nuclei through iron, 
are the main components of cosmic rays, and can be measured 
in the range to a PeV. 
The detailed mechanism and parameters governing propagation and acceleration of the galactic cosmic-rays will be investigated based on nuclei spectra measurements. 
CALET is expected to extend the limits of direct measurements.

\section{Instrument}
\begin{figure}[b!]
\begin{center}
\vspace*{-0.3cm}
\includegraphics[bb=0 0 722 227, width=\hsize]{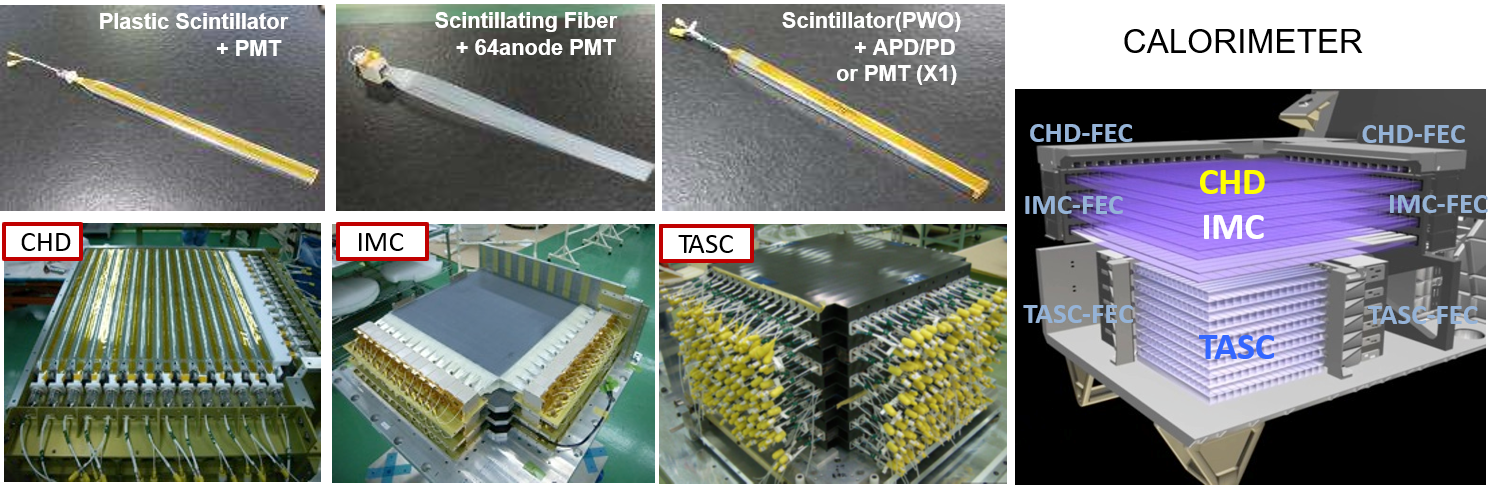}
\end{center}
\vspace*{-0.5cm}
\caption{CALET calorimeter consisting of three sub detectors, i.e., a CHarge Detector (CHD),
an IMaging Calorimeter (IMC), and Total AbSorption Calorimeter (TASC).
}
\label{fig:instrument}
\vspace*{-0.2cm}
\end{figure}
The CALET calorimeter (see the right-hand panel of Fig.~\ref{fig:instrument}) 
consists of a Charge Detector (CHD), which identifies the charge of 
the incident particle~\cite{pier2011,pier2013}, 
an IMaging Calorimeter (IMC), which reconstructs the track of 
the incident particle and 
records the initial shower development with fine resolution, 
and a Total AbSorption Calorimeter (TASC), which absorbs the entire 
energy of the incoming particle and identifies the particle species 
using hodoscopic lead-tungstate crystal arrays. 
The components and read-out sensors are summarized 
in the left panels of Fig.~\ref{fig:instrument}.

Combining these sub-detectors as well as the trigger system and data acquisition system, 
the CALET instrument 
features (1) a proton rejection factor larger than 10$^5$, 
(2) a 2\% energy resolution above 20~GeV for electrons, 
(3) very wide dynamic range from 1~GeV to 1~PeV,
(4) charge resolution of 0.1--0.3 electron charge unit from protons to above iron (up to $Z=40$), 
(5) an angular resolution of 0.1 to 0.5$^\circ$, 
and (6) a geometrical factor of the order of 0.1~m$^2$sr. 

\begin{figure}[bth!]
\begin{center}
\includegraphics[bb=0 0 697 393, width=0.8\hsize]{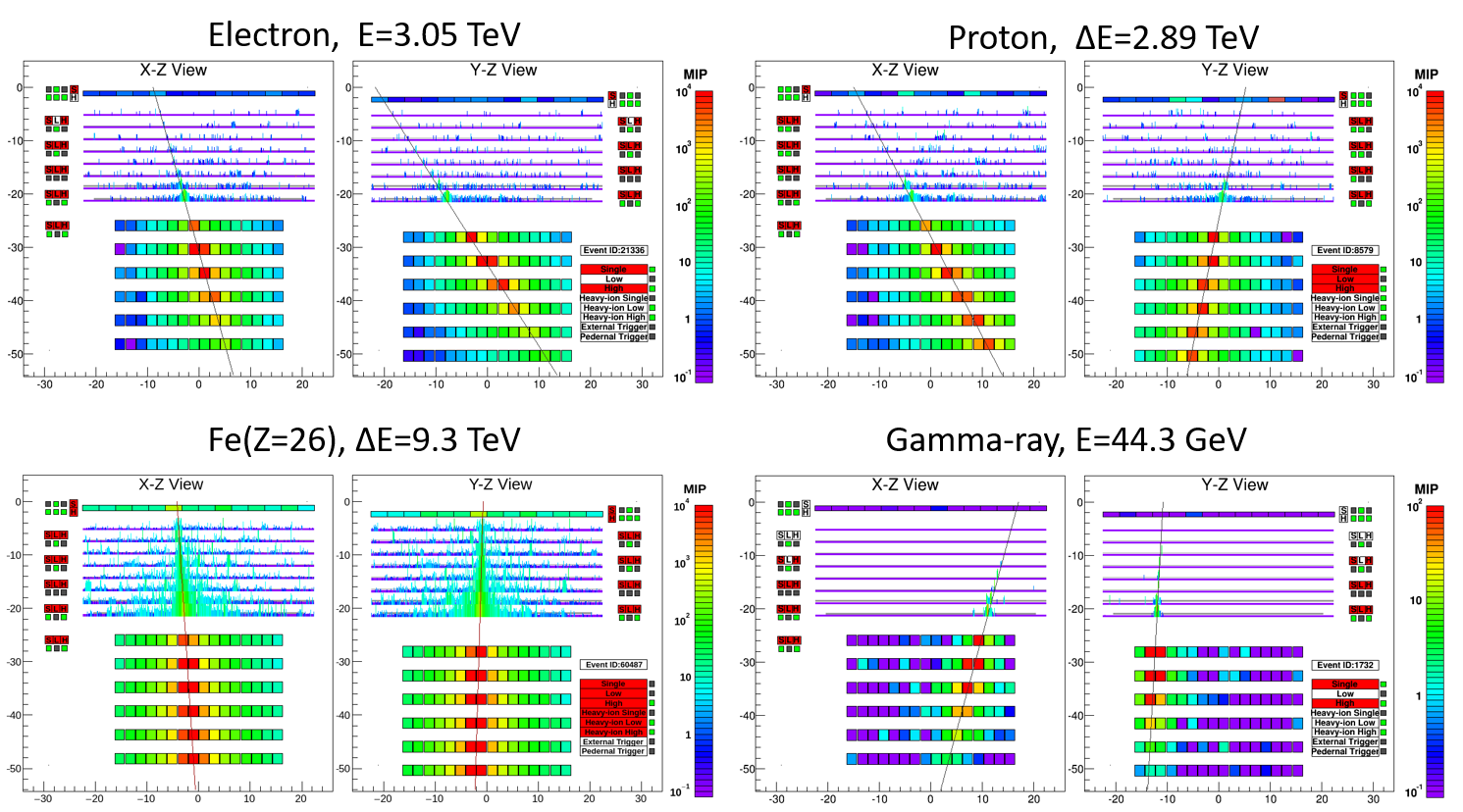}
\end{center}
\vspace*{-0.2cm}
\caption{Event examples of high-energy showers collected by CALET onboard the ISS.
({\it Top Left}) a 3~TeV electron candidate,
({\it Top Right}) a proton candidate with equivalent shower energy,
({\it Bottom Left}) an iron candidate with shower energy of 9.3~TeV, and
({\it Bottom Right}) a 44~GeV gamma-ray candidate.
}
\label{fig:events}
\vspace*{-0.3cm}
\vspace*{-0.3cm}
\end{figure}

Figure~\ref{fig:events} summarizes the CALET capability of particle identification.
Top left, top right, bottom left, and bottom right panels show 3~TeV electron candidate,
proton candidate with equivalent shower energy, iron candidate with shower energy of 9.3~TeV,
gamma-ray candidate with 44~GeV reconstructed energy, respectively.
The calorimeter with 30 radiation-length on-axis thickness absorbs the full electron shower energy 
even in the TeV range.
Charge measurement using CHD and IMC separates each of the elements from $Z=1$ to 26 and above.
Gamma-rays are identified as charge zero because they do not produce any signal before 
the pair creation.
While both of electrons and protons have $Z=1$, they can be separated using 
the differences in their shower shapes. 
Because of the continuing showering activity in the lower part of the TASC 
due to subsequent interactions of secondary pions, electrons, and protons are easily 
separated by a simple cut even in the TeV region. In addition to this, various parameters 
characterizing the shower shape can be utilized 
to improve the separation power~\cite{CALET2017,CALET2018}.

\section{On-Orbit Operations}
For on-orbit operation of CALET, JAXA Ground Support Equipment (JAXA-GSE) and the Waseda CALET Operations Center (WCOC) were set up at Tsukuba Space Center and Waseda University, respectively. The data taken by the CALET instrument are transferred from the ISS to JAXA using NASA's data relay system. The scientific operations of CALET are planned at WCOC~\cite{asaoka2018} based for example on the variation of the geomagnetic rigidity cutoff depending on ISS position. 
Accordingly is the observation mode of CALET controlled by scheduled command sequences.
These sequences define the time profile of calibration and data acquisition tasks as for example the 
recording pedestal and penetrating particle events, as well as switching on/off observation modes such as a low-energy electron trigger operating at high geomagnetic latitude, a low-energy gamma-ray (LE-$\gamma$) trigger operating at low geomagnetic latitude, and an almost continuously active ultra heavy trigger mode, during each ISS orbit. 
Maximum exposure to high-energy 
\begin{wrapfigure}{r}{7cm}
\vspace*{-0.8cm}
\begin{center}
\includegraphics[width=\hsize]{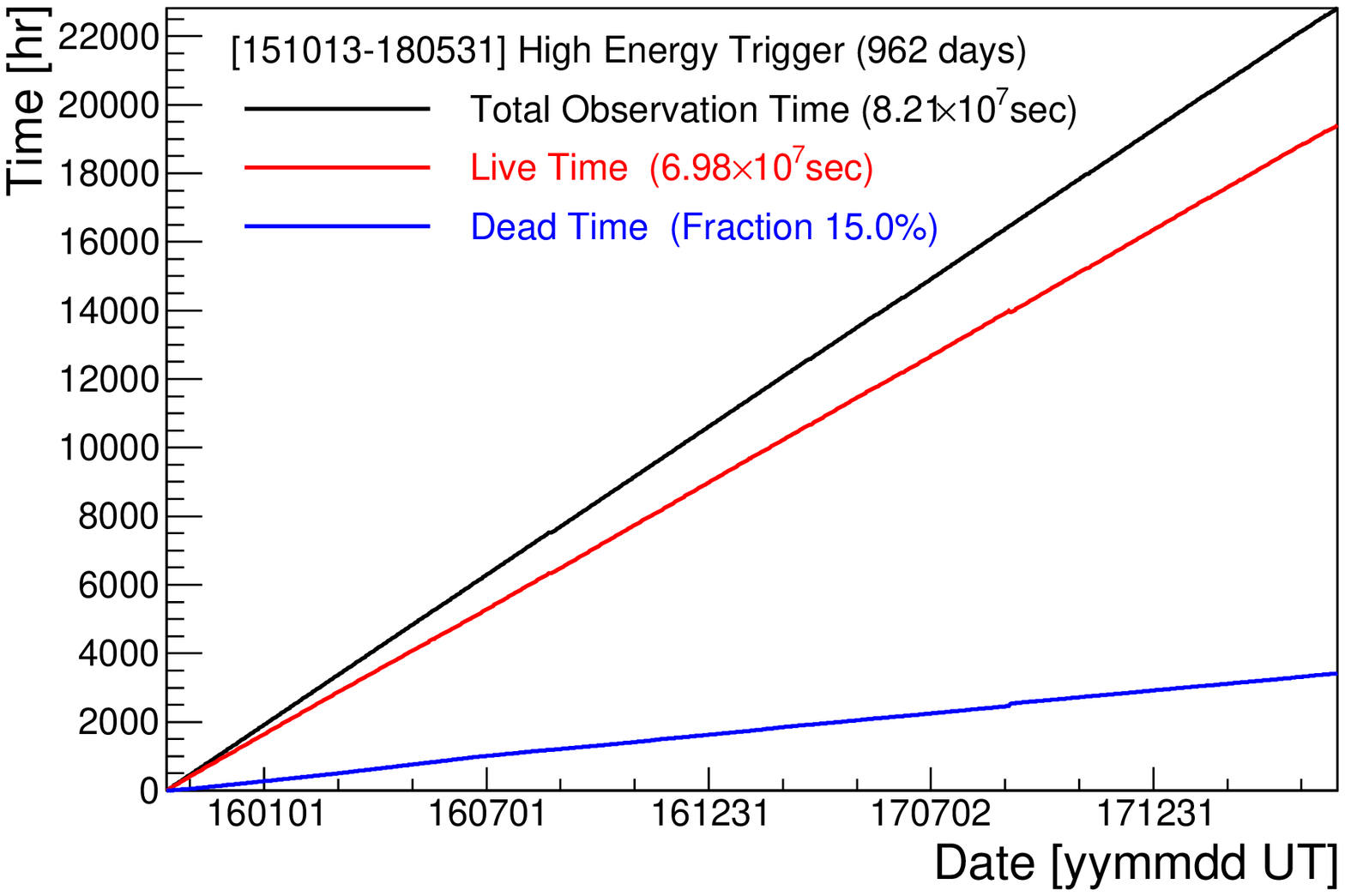}
\end{center}
\vspace*{-0.8cm}
\caption{
Cumulative of observation time (black line) for HE trigger. 
The red and blue lines indicate the live and dead times, respectively. 
}
\label{fig:stat}
\begin{center}
\includegraphics[width=0.9\linewidth]{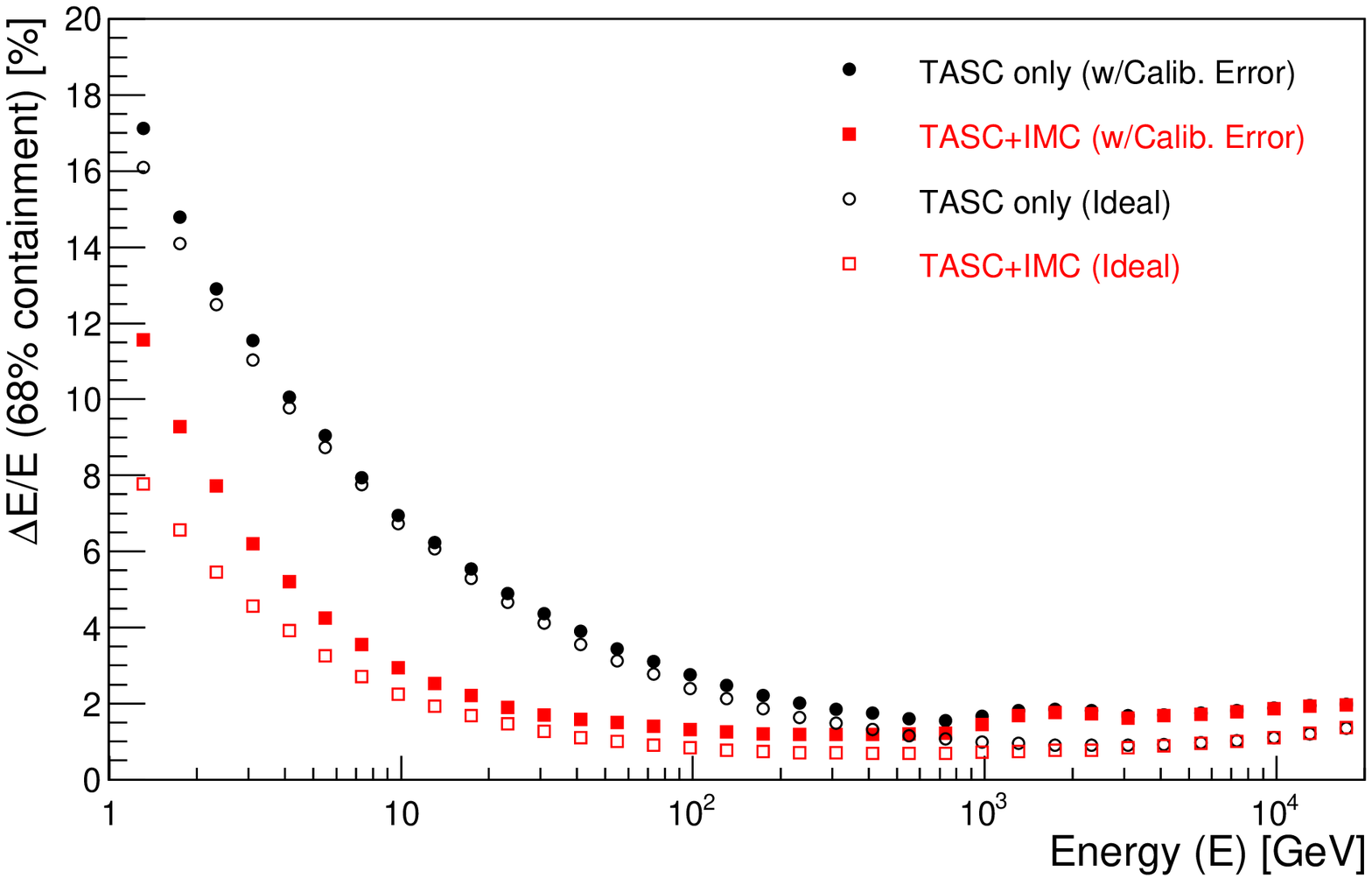}
\includegraphics[width=1.0\linewidth]{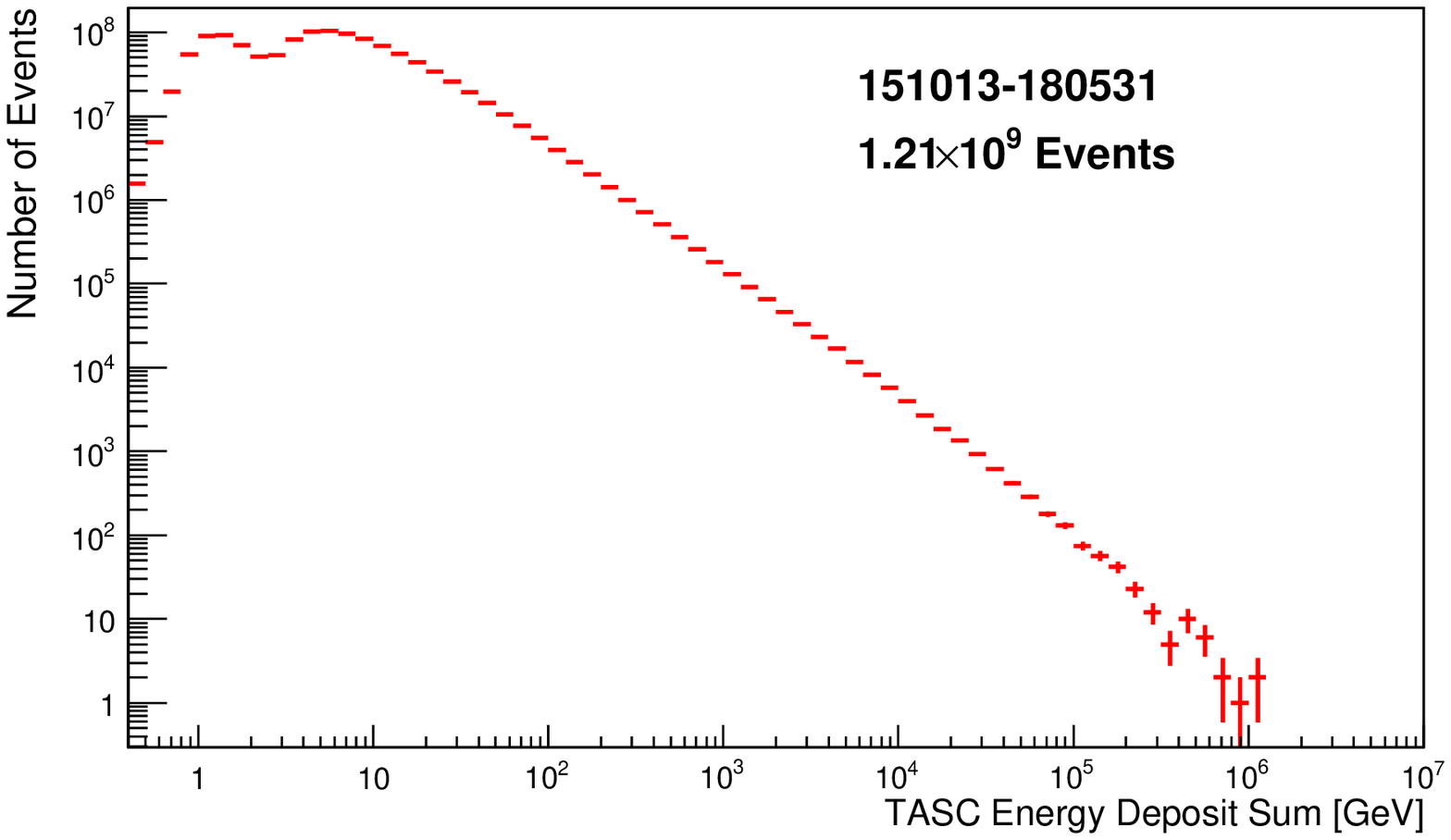}
\caption{({\it Top}) Energy resolution for electron measurements~\cite{asaoka2017},
({\it Bottom}) TASC energy deposit spectrum using all triggered events up to May, 2018.
\vspace*{-0.8cm}
}
\label{fig:calib}
\end{center}
\end{wrapfigure}
electrons and other high-energy shower events is ensured by an always active high-energy (HE) trigger mode.

As of May 31, 2018, the total observation time is 962~days with a live time fraction of $\sim$84\% relative to the total time. Close to 630~million events were taken with the HE (E $>$10~GeV) trigger mode. Figure~\ref{fig:stat} shows the accumulated live time 
for the HE trigger. 
Cumulative observation time has increased without significant interruption since scientific operation start in October 2015. Data transmission from JAXA-GSE to WCOC and data processing for scientific analysis at WCOC also proceeded smoothly.

\section{Calibration}
Energy calibration is a key procedure for CALET as a calorimeter instrument 
to achieve high precision and accurate measurements. 
While excellent energy resolution inside the TeV region is 
one of the most important features of a thick calorimeter instrument
like CALET, 
calibration errors must be carefully assessed 
and taken into account in the estimation of the actual energy resolution.

Our energy calibration~\cite{asaoka2017} includes the evaluation of the conversion factors
between ADC units and energy deposits, ensuring linearity over each 
gain range (TASC has four gain ranges for each channel), 
and provides a seamless transition between neighboring gain ranges. 
Temporal gain variations occurring during long time observations are 
also corrected for by the calibration procedure~\cite{CALET2017}. 

The errors at each calibration step, such as the correction of 
position and temperature dependence, consistency between energy deposit 
peaks of non-interacting protons and helium, linear fit error of each gain range, and 
gain ratio measurements, as well as slope extrapolation, are included 
in the estimation of the energy resolution.

As a result, a very high resolution of 2\% or better is achieved above 20~GeV~\cite{asaoka2017}
as shown in the top panel of Fig.~\ref{fig:calib}.
It should be noted that even with such a detailed calibration, the determining factor 
for the energy resolution is the calibration uncertainty, as the intrinsic resolution of 
CALET is $\sim$1\%. 
Intrinsic resolution refers to the detector's capability by design, taking advantage of
the thick, fully-active total absorption calorimeter.
Also important is the fact that the calibration error in the lower gain ranges is crucial
for the spectrum measurements in the TeV range.

The bottom panel of Fig.~\ref{fig:calib} shows the TASC energy deposit spectrum 
using all triggered events through the end of May, 2018. 
The first bump is due to low-energy triggered events, while the second bump 
is caused by high-energy triggered events and the tail 
at high energy reflects the power-law nature of the cosmic-ray spectrum. 
The spectrum spans more than six orders of magnitude in energy with highest 
energy past a PeV, and the lowest energy below 1~GeV. 
This clearly demonstrates the CALET capability to observe 
cosmic rays over a very wide dynamic range.

\section{Results}

\subsection{All-Electron Spectrum}
A precise measurement of the all-electron (electron $+$ positron) 
spectrum in the TeV region might reveal interesting spectral features 
to provide the first experimental evidence of the presence of 
a nearby cosmic-ray source~\cite{nishimura1980, kobayashi2004}.
In addition, the unexpected increase of the positron fraction 
above 10~GeV established by PAMELA \cite{PAMELA-pe} and AMS-02 \cite{AMS02-pe} 
may require a primary source component for positrons in addition to 
the generally accepted secondary origin. 
Candidates for such primary sources range from 
astrophysical (pulsar) to exotic (dark matter).
Since these primary sources emit electron-positron pairs, 
it is expected that the all-electron 
spectrum would exhibit a spectral feature, near the 
highest energy range of the primary component.

The CALET collaboration published its first result on electrons
in the energy range from 10~GeV to 3~TeV~\cite{CALET2017}.
Subsequently, the DArk Matter Particle Explorer (DAMPE) collaboration published
their all-electron spectrum in the energy range from 25~GeV to 4.6~TeV~\cite{DAMPE2017}.
The latter publication was followed by many papers speculating about
the origin of a peak-like structure near 1.4~TeV in the DAMPE data.

\begin{figure}[bth!]
\vspace*{-1.0cm}
\begin{center}
\includegraphics[width=0.85\hsize]{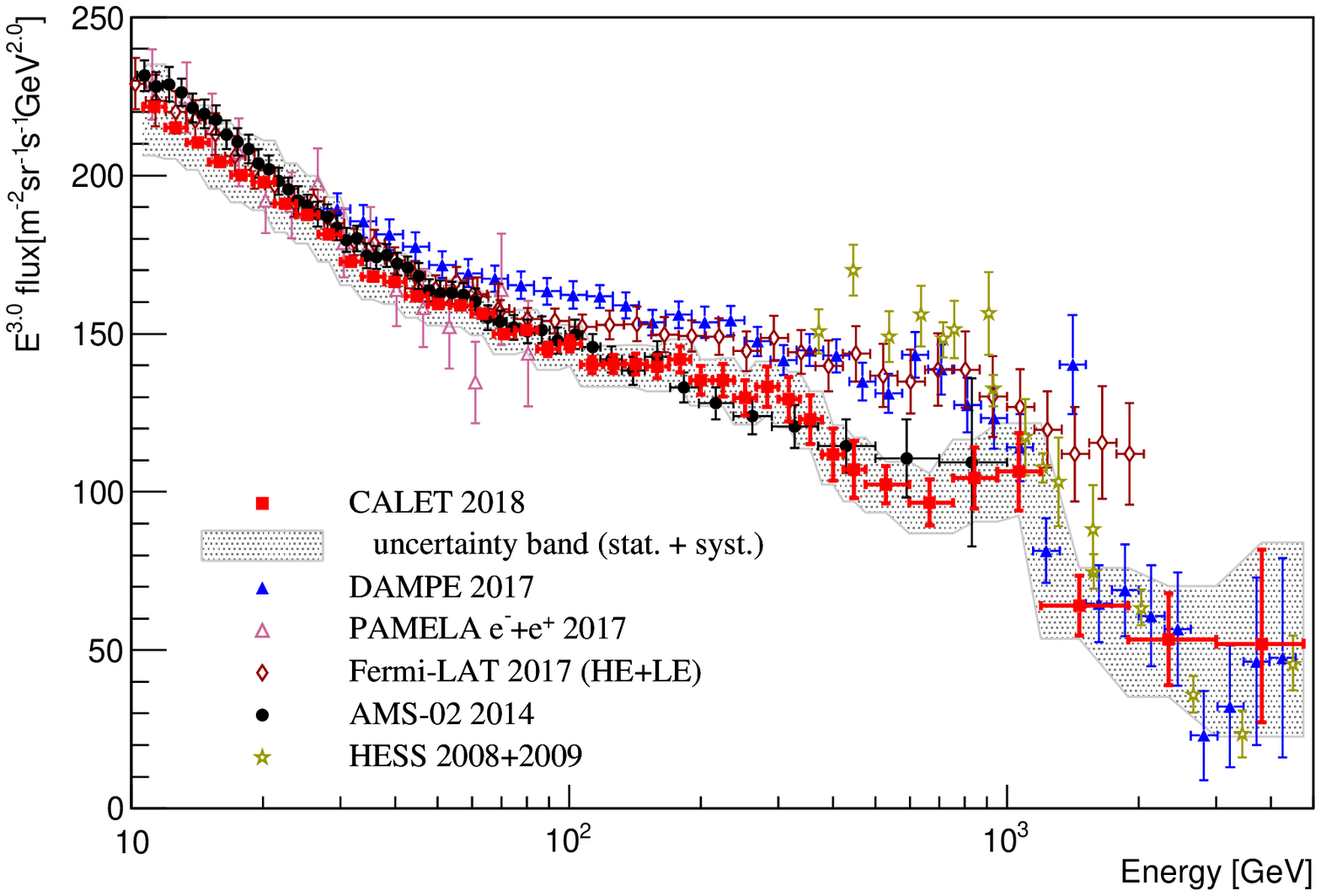}
\vspace*{-0.2cm}
\caption{Cosmic-ray all-electron spectrum measured by CALET 
from 10.6~GeV to 4.75~TeV~\cite{CALET2018},
where the gray band indicates the quadratic sum of 
statistical and systematic errors 
(not including the uncertainty on the energy scale). 
Also plotted are direct measurements 
in space~\cite{DAMPE2017,Pamela-e,Fermi2017-e,AMS02-e} and 
from ground-based experiments~\cite{HESS2008,HESS2009}. 
}
\label{fig:binCALET}
\end{center}
\vspace*{-0.3cm}
\begin{minipage}{0.49\hsize}
\begin{center}
\includegraphics[width=\hsize]{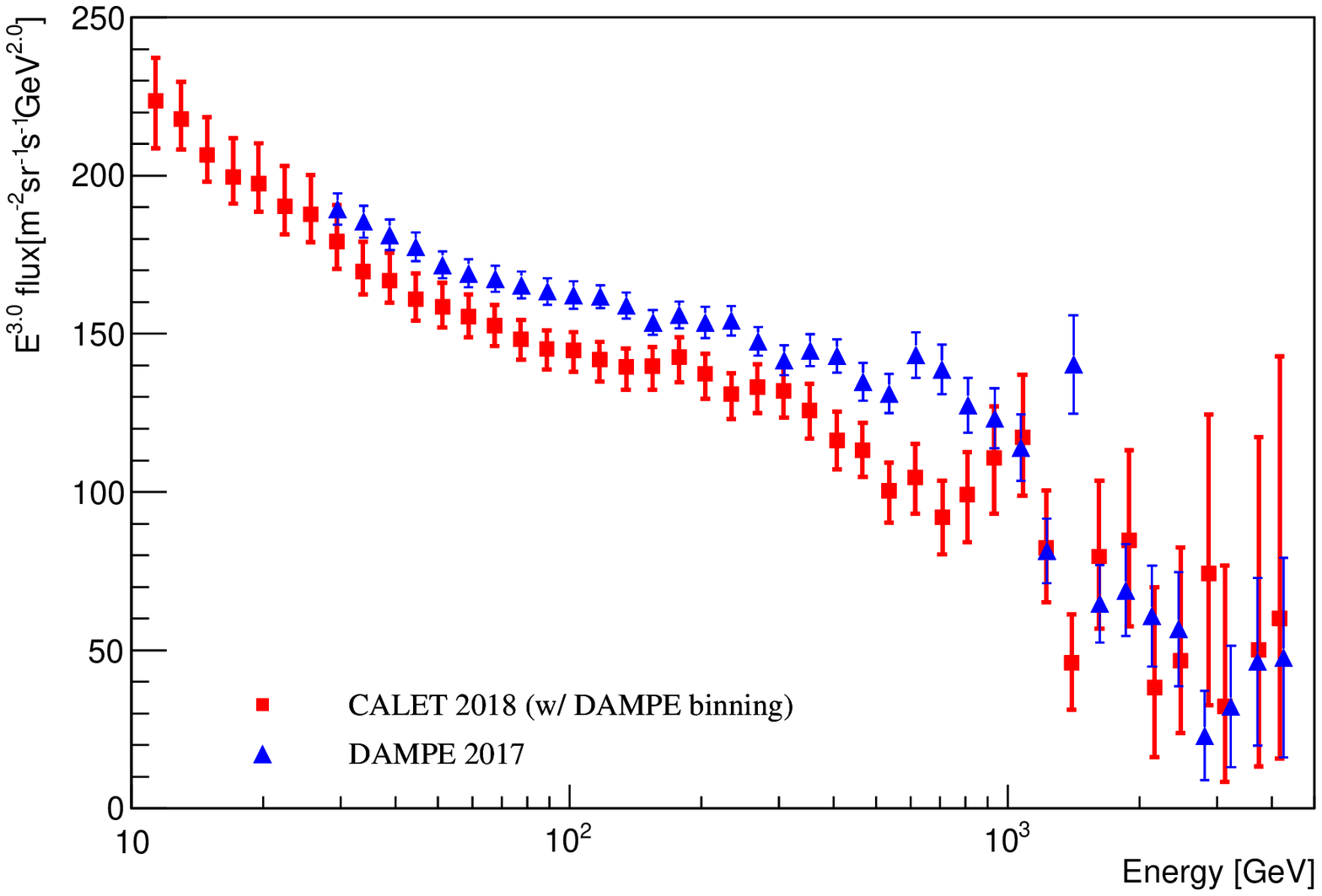}
\end{center}
\end{minipage}
\begin{minipage}{0.49\hsize}
\begin{center}
\includegraphics[width=\hsize]{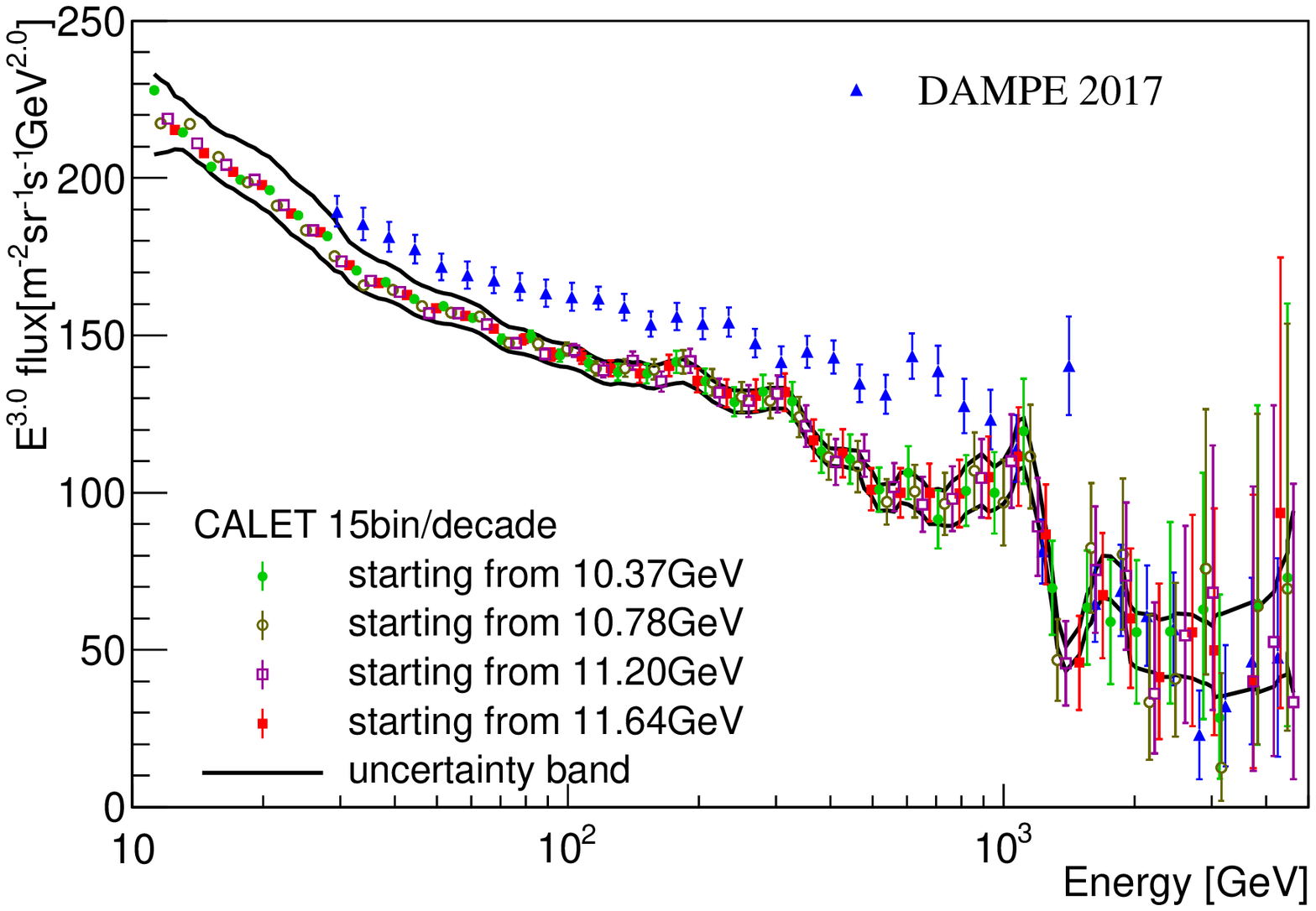}
\end{center}
\end{minipage}
\vspace*{-0.2cm}
\caption{
({\it Left}) Cosmic-ray all-electron spectrum measured by CALET from 10.6~GeV to 4.57~TeV~\cite{CALET2018},
using the same energy binning as DAMPE's result~\cite{DAMPE2017} and compared with it. 
The error bars indicate the quadratic sum of 
statistical and systematic errors.
({\it Right}) 
Study of possible binning related effects in the CALET all-electron spectrum~\cite{CALET2018}. 
}
\label{fig:binDAMPE}
\vspace*{-0.3cm}
\end{figure}

Recently, an updated version of the CALET all-electron spectrum
using 780 days of flight data and the full geometrical acceptance was published 
in the energy range from 11~GeV to 4.8~TeV~\cite{CALET2018}.
Figure~\ref{fig:binCALET} shows the updated all-electron spectrum obtained with CALET 
using the same energy binning as in our previous publication~\cite{CALET2017}, except for adding one extra bin at the
high energy end.
The error bars along horizontal and vertical axes indicate bin width 
and statistical errors, respectively. 
The gray band is representative of the quadratic sum of statistic and systematic errors.
The exhaustive study on the systematic uncertainties was performed and is described in Refs.~\cite{CALET2017, CALET2018} 
including their Supplemental Materials.

There are four important implications from the current status on the all-electron spectrum measurements.
First, CALET's spectrum is consistent with AMS-02 below 1~TeV. 
As both detectors have enough capability to identify electrons up to 1~TeV, and 
the detection principles are largely different (calorimeter versus magnet spectrometer),
the agreement is an important indication.
Secondly, 
there are two group of measurements: AMS-02 $+$ CALET vs Fermi$-$LAT $+$ DAMPE, 
indicating the presence of unknown systematic errors.
Thirdly, 
CALET observes flux suppression consistent with DAMPE within errors above 1~TeV.
No peak-like structure was found at 1.4~TeV in CALET data, irrespective of energy binning 
as shown in Fig.~\ref{fig:binDAMPE}.

In the left-hand plot of Fig.~\ref{fig:binDAMPE}, 
we have adopted exactly the same energy binning as DAMPE to show our spectrum.
The flux is inconsistent between the two experiments with a 4$~\sigma$ significance; 
the CALET data does not show any significant excess in the 1.4~TeV bin.
The significance includes the systematic errors quoted from both experiments.
Furthermore, possible binning related effects in the CALET all-electron spectrum
are studied with a shift of one fourth of the bin width as shown 
in the right-hand plot of Fig.~\ref{fig:binDAMPE}.
The solid curves in the figure show the energy dependent systematic uncertainty band.
The deviation due to binning is well 
below our energy dependent systematic uncertainty or statistical fluctuations.
Therefore, bin-to-bin migration and related effects are negligible compared to 
our estimated systematic uncertainties, 
in accordance with the estimated CALET energy resolution of 2\% above 20~GeV. 

\subsection{Hadrons}
Direct measurements of the high-energy spectra of 
each species of cosmic-ray nuclei up to the PeV
energy
scale provide 
information complementing all-electron observations and provides more detailed insight into the general conditions of cosmic-ray acceleration and propagation.
A possibly charge-dependent cutoff in the nuclei spectra is hypothesized to explain the ``knee'' in the all-particle spectrum. This hypothesis could only be investigated by a space experiment with sufficient exposure. 
The acceleration limit of supernova remnants calculated with nominal parameters is typically found to be far smaller than the energy of the ``knee''~\cite{ARBell2013} observed indirectly by ground detectors. Therefore, precise direct observation of the proton and helium spectra up to PeV energy is highly important. 
Also the spectral hardening observed in spectra of various nuclei calls for an careful investigation, with CALET's wide dynamic range from GeV to PeV energies allowing to study the feature unaffected by systematics from combination of spectra measured by different experiments.
Another detailed study will be done 
on the spectral behavior of heavier elements, including secondary-to-primary ratios up to 1 TeV/n energy region, which should yield important information about propagation parameters such as the diffusion coefficient.

\begin{wrapfigure}{r}{7cm}
\vspace*{-1.1cm}
\begin{center}
\includegraphics[bb=0 0 360 250, width=1.0\linewidth]{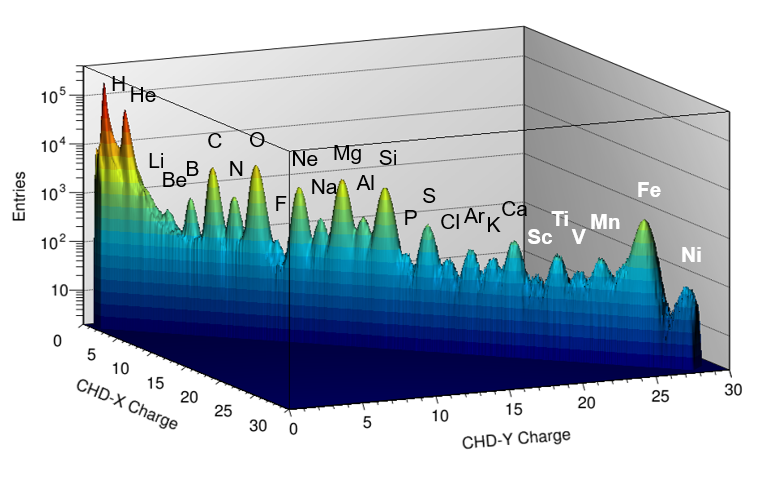}
\vspace*{-0.3cm}
\caption{CHD charge separation.
}
\label{fig:chd_charge}
\end{center}
\vspace*{-0.8cm}
\end{wrapfigure}
Figure~\ref{fig:chd_charge} illustrates 
the charge identification capability of CALET based on the CHD data only, showing clear separation of protons, helium and nuclei up to iron and nickel. The IMC provides an independent and about equally powerful charge separation capability for light elements.  Until now preliminary results on protons~\cite{pier2017} and heavier nuclei~\cite{akaike2017,akaike2018} have been presented, taking advantage of the accurate charge determination and wide energy range. 
The first results on the nuclei spectra will be published soon, 
including a detailed assessment of systematic uncertainties. 

\subsection{Gamma-Rays}
With a fully active 30 radiation-length thick calorimeter, 
CALET is capable of measuring gamma-rays up to the TeV region.
In addition to HE trigger, CALET uses LE-$\gamma$ 
trigger to be sensitive to gamma rays with primary energies down to 1~GeV.
To avoid large dead-time fraction, however, LE-$\gamma$ trigger is activated only  
at low geomagnetic latitudes or following gamma-ray 
burst triggered onboard by CALET gamma-ray bust monitor (CGBM).

\begin{figure}[bth!]
\begin{center}
\begin{minipage}{0.49\hsize}
\begin{center}
\includegraphics[bb=0 0 1176 588, width=\hsize]{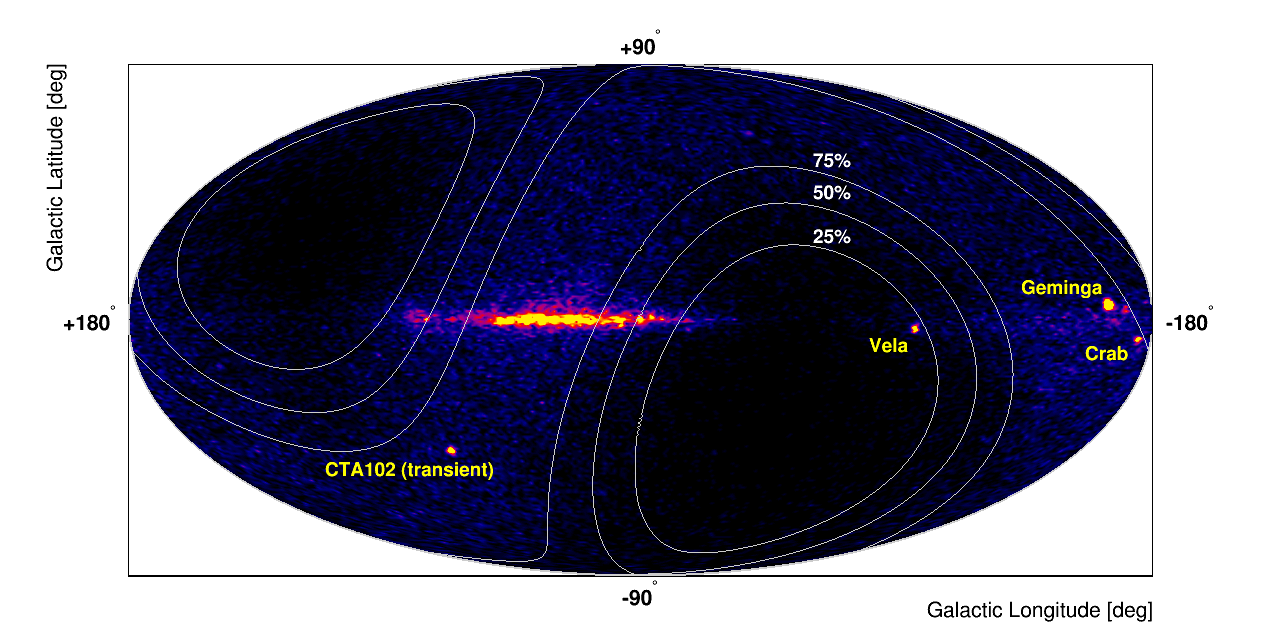}
\end{center}
\end{minipage}
\begin{minipage}{0.49\hsize}
\begin{center}
\includegraphics[bb=0 0 520 245, width=1.0\hsize]{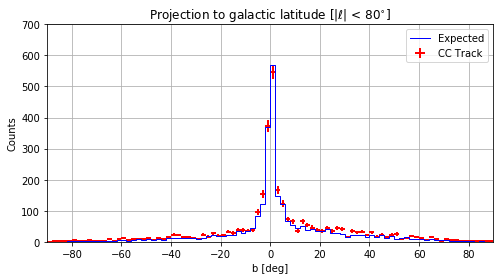}
\end{center}
\end{minipage}
\caption{
({\it Left}) Gamma-ray sky map shown in a Mollweide projection of galactic coordinates.
White contours show the relative level of exposure compared to the maximum on the sky.
({\it Right}) Projection of the observed and expected number of photons onto 
galactic latitude for the galactic plane region $|l| < 80^o$ for the energy range from 1 to 100~GeV.
}
\vspace*{-0.5cm}
\label{fig:gamskyCC}
\end{center}
\end{figure}

The first 24 months of on-orbit scientific data provide
valuable characterization of the performance of the calorimeter 
based on the analyses of the gamma-ray data set~\cite{GI-CALET2018}. 
It includes optimization of event selection criteria, 
calculation of effective area, 
determination of point spread function, 
confirmation of absolute pointing accuracy, 
observation of bright point sources and study of diffuse components.
Based on the developed analysis method, 
CALET gamma-ray sky seen by LE-$\gamma$ trigger is shown in the left-hand panel of Fig.~\ref{fig:gamskyCC},
where galactic emission and bright gamma-ray sources are clearly identified.
In the right-hand panel of Fig~\ref{fig:gamskyCC}, projection of the observed 
and expected number of photons onto galactic latitude for the galactic plane 
region $|l| < 80^o$ is shown. Expected number of photons are calculated using 
Fermi/LAT flux map and CALET's exposure in the same region of the sky.
The very good consistency confirmed our sensitivity.
It should be noted, that it is important to take the effects of ISS structures in 
the field-of-view into account. 

Gamma-ray transients are also an important observational target for CALET.
CGBM detected nearly 60 GRBs ($\sim$20\% short GRB among them) 
per year in the energy range of 7~keV--20~MeV, as expected~\cite{yamaoka2017}.  
To search for GeV-energy counterpart emission from such 
sources also detected by other instruments,
we check the CAL data at the reported trigger times for
gamma-ray candidates. 
For events checked 
based on CGBM, 
Swift, and Fermi/GBM triggers, 
no significant counterparts have been detected at this stage 
for timescales ranging from 1~s to 1~hr~\cite{GI-CALET2018}.
Regarding the counterpart search for gravitational wave events,
combined analyses 
of CGBM and calorimeter were performed for GW151226,
resulting in our upper limits set on X-ray and gamma-ray counterparts~\cite{GW-CALET2016}.
Furthermore, complete search results of CALET calorimeter on 
the LIGO/Virgo's Observation Run~2 has been published recently~\cite{GW-CALET2018}. 

\section{Summary and Prospects}
CALET was successfully launched on Aug.~19, 2015, 
and 
detector performance for scientific observation has been continuously very stable 
since Oct.~13, 2015. 
Careful calibrations using ``MIP'' signals of the non-interacting protons and helium events 
have been successfully carried out, 
and the linearity of the energy measurements up to 10$^6$~MIPs 
was 
established based on observed events~\cite{asaoka2018,asaoka2017}.

\begin{wrapfigure}{r}{8.0cm}
\vspace*{-1.0cm}
\begin{center}
\includegraphics[width=1.08\linewidth]{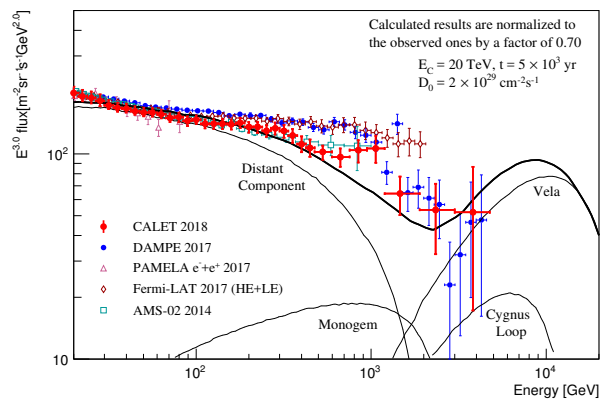}
\vspace*{-0.5cm}
\caption{Current situation of all-electron spectrum directly measured in space, 
together with the model calculation~\cite{kobayashi2004}, 
where main model parameters are shown in the plot. 
}
\label{fig:alle_prospects}
\end{center}
\vspace*{-0.3cm}
\end{wrapfigure}
The all-electron spectrum~\cite{CALET2017} has been published with extended energy range from 11~GeV to 4.8~TeV~\cite{CALET2018} and improved statistics. Figure~\ref{fig:alle_prospects} presents the current situation of all-electron spectrum direct measurements in space. Five years or more of observations with CALET will triple the statistics, 
which together with a reduction of systematic errors based on a better understanding 
of the detector with 
an increased amount of the flight data will lead to significantly improved precision. This will allow a refined study of the possible fine structures around a few hundred GeV and $\sim$1~TeV, which are currently not significant, which might shed light on the origin of the positron excess.
An extension of the high-energy reach using improved statistics and electron separation methods might for the first time reveal a charged cosmic-ray signature of a local accelerator. 

Preliminary results on protons~\cite{pier2017}, primary and secondary
nuclei up to $Z=26$ and their ratios (for example, boron to carbon)~\cite{akaike2017,akaike2018} were presented so far, demonstrating CALET's wide dynamic range of energy measurement from 1~GeV to 1~PeV and the accuracy of its charge determination capability. Their publication is foreseen in the near future, and will address important questions in cosmic-ray physics, 
such as the charge dependence of the acceleration limit in supernova remnants, the universality of the widely observed spectral hardening, and the energy dependence of the diffusion coefficient.
The relative abundance of the ultra heavy nuclei up to $Z=40$ is also analyzed~\cite{brian2017}.

There is also significant progress in CALET's gamma-ray analysis. Based on the science data taken in two years on orbit, the performance of the gamma-ray 
measurements 
has been characterized~\cite{GI-CALET2018}. These results confirm the capability of CALET to observe gamma rays in the energy range from $\sim$1~GeV to over 100~GeV. 
CALET's current results on  electromagnetic counterpart search for gravitational wave events~\cite{GW-CALET2016,GW-CALET2018} show the great potential of follow-up observations during the upcoming LIGO/Virgo's third observation run (Observation Run~3). The continuous GeV gamma-ray sky observation with CALET complements the coverage by other missions and may help to identify unexplored high-energy emissions from future transient events. Watching for various transient phenomena including those in gamma rays, but not limited to them is an important task for CALET 
as an on-orbit observation strategy. 
Through the detection of many events of MeV electrons originating from the radiation belt~\cite{Kataoka2016},  a phenomenon called relativistic electron precipitation, space weather was added as an additional observational target for CALET 
after the start of on-orbit operations. 

The so far excellent performance of CALET and the outstanding quality of 
the data suggest that a 5-year (or more) observation period 
will most likely provide a wealth of interesting new results.

\section*{Acknowledgment}
We gratefully acknowledge JAXA's contributions to the development of CALET and to the
operations onboard the International Space Station. 
We also wish to express our sincere gratitude to ASI and NASA for
their support of the CALET project. This work was supported in part by 
JSPS Grant-in-Aid for Scientific Research (S) (No.~26220708), 
JSPS Grant-in-Aid for Scientific Research (B) Number~17H02901, 
JSPS Grant-in-Aid for Scientific Research (C) Number~16K05382,
and by the MEXT-Supported Program for the Strategic Research Foundation 
at Private Universities (2011-2015) (No.~S1101021) at Waseda University.
The CALET effort in the United States is supported by NASA through 
Grants No.~NNX16AB99G, No.~NNX16AC02G, and No.~NNH14ZDA001N-APRA-0075.

\section*{References}
\providecommand{\noopsort}[1]{}\providecommand{\singleletter}[1]{#1}%
\providecommand{\newblock}{}

\end{document}